\documentclass{emulateapj} 
\usepackage{natbib}
\usepackage{epsfig}
\usepackage{graphicx} 
\usepackage{subfigure}
\usepackage{float}
\usepackage{amsmath}
\usepackage{color}
\usepackage{amssymb}
\usepackage{amsfonts}
\usepackage{units}
\usepackage{mathtools}
\usepackage{bm}
\usepackage[colorlinks,linkcolor=blue,anchorcolor=green,citecolor=blue]{hyperref}
\def\be{\begin{eqnarray}} 
\def\ee{\end{eqnarray}}

\shorttitle{Fast optical bursts associated with FRBs}
\shortauthors{Yang, Zhang \& Wei} 
\begin{document} 

\title{How bright are fast optical bursts associated with fast radio bursts?}

\author{Yuan-Pei Yang\altaffilmark{1}, Bing Zhang\altaffilmark{2} and Jian-Yan Wei\altaffilmark{3}} 

\affil{$^1$South-Western Institute for Astronomy Research, Yunnan University, Kunming, Yunnan, P.R.China; yypspore@gmail.com;\\
$^2$ Department of Physics and Astronomy, University of Nevada, Las Vegas, NV 89154, USA; zhang@physics.unlv.edu \\
$^3$ National Astronomical Observatories, Chinese Academy of Sciences, Beijing 100012, China}

\begin{abstract}
The origin of fast radio bursts (FRBs) is still unknown. Multi-wavelength observations during or shortly after the FRB phase would be essential to identify the counterpart of an FRB and to constrain its progenitor and environment. In this work, we investigate the brightness of the ``fast optical bursts'' (FOBs) associated with FRBs and the prospects of detecting them. We investigate several inverse Compton (IC) scattering processes that might produce an FOB, including both the one-zone and two-zone models. We also investigate the extension of the same mechanism of FRB emission to the optical band. We find that a detectable FOB with the current and forthcoming telescopes is possible under the IC scenarios with very special conditions. In particular, the FRB environment would need to invoke a neutron star with an extremely strong magnetic field and an extremely fast spin, or an extremely young supernova remnant (SNR) surrounding the FRB source. Furthermore, most electrons in the source are also required to have a fine-tuned energy distribution such that most of the IC energy is channeled in the optical band. We conclude that the prospect of detecting FOBs associated with FRBs is low. On the other hand, if FOBs are detected from a small fraction of FRBs, these FOBs would reveal extreme physical conditions in the FRB environments. 
\end{abstract}


\section{Introduction}\label{sec1}

Fast radio bursts (FRBs) are mysterious millisecond-duration radio transients, which are characterized by an excess dispersion measure (DM) respect to the Galactic value, high peak flux, and an all-sky distribution \citep{lor07,tho13,cha17,ami19a,ami19b}. 
Growing evidence suggests that FRBs are of a cosmological origin: 1. The repeating source, FRB 121102, has been located in a dwarf galaxy at $z=0.19273$ \citep{cha17,mar17,ten17}; 2. Other non-repeating FRBs have an all-sky distribution and DM excess with respect to the Galactic contribution \citep{tho13}; 3. The dispersion-brightness relation from a wide-field survey (ASKAP) \citep{sha18} shows that the excess DM of FRBs can be used as a proxy for cosmological distances.

Thanks to the multi-wavelength follow-up observations, a persistent radio counterpart with a continuous spectrum peaking at a few $\unit{GHz}$ is found to be associated with FRB 121102 \citep{cha17,mar17,ten17}. However, the localization of a non-repeating FRB with a low DM, FRB 171020, suggested that not all FRBs have an associated persistent radio counterpart \citep{mah18}. 

So far, there is no confirmed multi-wavelength or multi-messenger transient being associated with any FRB\footnote{Some putative counterparts have been reported, e.g. a bright radio variable source (likely an AGN) potentially associated with FRB 150418 \citep{kea16,wil16}, and a sub-threshold long GRB potentially associated with FRB 131104 \citep{del16}, but the associations are not confirmed due to the large error circle of FRBs. } \citep[e.g.][]{pet15,cal16,zha17,aar18,eft18,mag18}.
In general, the following two physical mechanisms can give rise to an FRB-associated high-frequency counterpart during the FRB phase: 1. the inverse Compton (IC) scattering processes associated with an FRB, either for the one-zone case for which the IC and FRB emission processes are from the same emission region (the electrons responsible for the FRB emission and the IC emission could be same or different) or for the two-zone case for which the IC process is located at a different region from the FRB emission; 2. the extension of the same mechanism of FRB emission to higher frequencies.
These counterparts are generic and weakly dependent on the origin of FRBs\footnote{Besides the above two cases, the afterglow associated with the FRB explosion energy is also weakly dependent on the FRB origin. However, the FRB afterglow has a much longer duration than that of the FRB itself \citep{yi14}.}.

In this work, we discuss the possible optical flashes temporally associated with the FRBs, which we call ``fast optical bursts'' or FOBs\footnote{Such events were speculated by \cite{lyu16} without physical modeling.}. We first generally discuss the detectable flux of FOBs with the current and upcoming telescopes in Section \ref{sec2}. We then investigate the possible IC scattering processes that may produce an FOB in Section \ref{sec3}. The FOBs generated via the same intrinsic mechanism of FRBs are discussed in Section \ref{sec4}. The results are summarized in Section \ref{sec5} with some discussions.

\section{Detectability of FOBs} \label{sec2}

In order to more straightforwardly compare the predicted FOB peak fluxes with the instrumental sensitivity, we first discuss the minimum detectable source fluxes of FOBs that are related to the limiting magnitudes of the current and forthcoming transient optical telescopes.
The duration of an FOB could be shorter than the telescope exposure time (which is typically $\gtrsim$ tens of seconds), which causes the observed effective flux less than the intrinsic peak flux. Let us assume that the FOB has a peak flux $F_\nu$ and a duration $\tau$, then the observed effective flux may be estimated as (noise level ignored)
$F_{\nu,{\rm eff}}\sim\min(\tau/T,1)F_\nu$, where $T$ is the telescope exposure time. In the optical band, the magnitude of a source is related to its flux through $m=-2.5\log_{10}\left(F_{\nu}/3631~\unit{Jy}\right)$, or  \citep{lyu16}
\be
m=
20.8-2.5\log_{10}\left(\frac{\tau_{\rm ms}F_{\nu,{\rm Jy}}}{T_{60}}\right)
\ee
for $\tau\lesssim T$, where $F_{\nu,{\rm Jy}}$ is the peak flux in Jansky, $\tau_{\rm ms}$ is the optical pulse duration in millisecond, and $T_{60}$ is the exposure time normalized to $60~\unit{s}$. Therefore, for a telescope with a limiting magnitude $m_\ast$, the limiting intrinsic flux of an FOB would be 
\be
F_{\nu \ast}=
\left(\frac{T_{60}}{\tau_{\rm ms}}\right)10^{(8.32-0.4m_\ast)}~\unit{Jy}
\ee
for $\tau\lesssim T$.
The Palomar Transient Factory (PTF) can reach a magnitude $m_\ast=20$ during a $60~\unit{s}$ exposure time \citep{law09}. Therefore, the limiting flux needs to be $F_{\nu \ast}\simeq2~\unit{Jy}$ for $\tau\sim1~\unit{ms}$ and $T\sim60~\unit{s}$. For an optical transient with $\tau\gtrsim T$, one has $F_{\nu \ast}\simeq4\times10^{-5}~\unit{Jy}$. The forthcoming Large Synoptic Survey Telescope (LSST) can reach a magnitude $m_\ast=25$ within two exposures of $\sim15~\unit{s}$ each \citep{ive08}. Thus, the limiting flux is $F_{\nu \ast}\simeq0.01~\unit{Jy}$ for $\tau\sim1~\unit{ms}$ and $T\sim30~\unit{s}$. For $\tau\gtrsim T\sim30~\unit{s}$, one has $F_{\nu \ast}\simeq4\times10^{-7}~\unit{Jy}$. In the following discussion, we take a limiting intrinsic flux $F_{\nu \ast}\simeq0.01~\unit{Jy}$ for $\tau\sim1~\unit{ms}$ and $F_{\nu \ast}\simeq10^{-7}~\unit{Jy}$ for $\tau\gtrsim T\sim\text{a few}\times10~\unit{s}$ as an optimistic minimum detectable flux and compare various model predictions with this value. 

\section{Inverse Compton scattering}\label{sec3}

\subsection{A generic model}\label{ic}
 
\begin{figure}
\centering
\includegraphics[angle=0,scale=0.3]{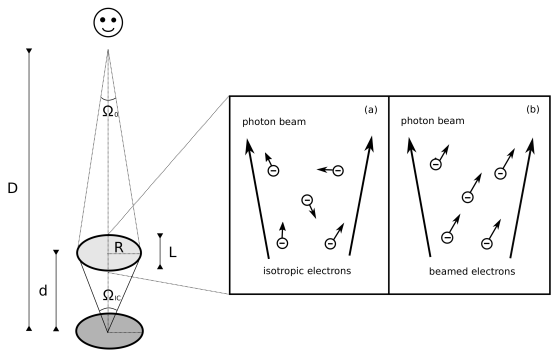}
\caption{A cartoon figure for IC scattering of FRB emission. The dark grey area denotes the seed photon (FRB) emission region, and the light grey area denotes the IC scattering region. Panel (a) shows the IC scattering process by isotropic electrons, and Panel (b) shows the IC scattering process by beamed electrons.}\label{fig1}
\end{figure}

In this section, we discuss a model-independent IC scattering process\footnote{For radio sources with high brightness temperatures, e.g., pulsars and FRBs, the induced (or stimulated) Compton scattering process may be important \citep[e.g.][]{wil78,lu18}. However, induced Compton scattering is important only when the frequency change before and after the scattering is small, e.g., $\Delta\nu\lesssim\nu$, which causes the suppression of the radio flux. For the IC process producing higher-frequency emission, the spontaneous scattering dominates, and the effect of induced Compton scattering is not important. }.
In the following discussion, we assume that the IC scattering region is optically thin. For a single electron with Lorentz factor $\gamma\gg1$, the emission power is 
\be
P_{\rm compt}=\frac{4}{3}\sigma_{\rm T}c\gamma^2U_{\rm ph}, 
\ee
where $U_{\rm ph}$ is the seed photon energy density. Consider a beam of electrons with density $n_e$ and Lorentz factor $\gamma$. The emission coefficient at the peak frequency is approximately
\be
j_{\nu,{\rm IC}}\sim\frac{n_eP_{\rm compt}}{\nu\Delta\Omega}\sim\frac{4}{3}hc\sigma_{\rm T}\frac{n_en_{\rm ph}}{\Delta\Omega},~~~{\rm for}~\nu\sim\gamma^2\nu_0, \label{emission}
\ee
where $\nu_0$ is the typical frequency of the incident photon (i.e., the FRB emission), and $n_{\rm ph}\sim U_{\rm ph}/h\nu_0$ is the number density of the incident photons. 

Next, we consider the effect of electron momentum distribution. Define the geometric beaming solid angle of the electron stream as $\Delta \Omega_e$. 
For a single electron with Lorentz factor $\gamma$, the beaming solid angle of the radiation is $\sim\pi/\gamma^2$ \citep{ryb79}.
If $\Delta\Omega_e<\pi/\gamma^2$, the radiation beaming solid angle is $\Delta\Omega\sim\pi/\gamma^2>\Delta\Omega_e$. On the other hand, if $\Delta\Omega_e>\pi/\gamma^2$, the radiation beaming solid angle is approximately equal to the electron beaming solid angle itself, i.e., $\Delta\Omega\sim\Delta\Omega_e$.
In general, the radiation beaming solid angle can be written as
\be
\Delta\Omega\simeq\max\left(\Delta\Omega_e,\frac{\pi}{\gamma^2}\right)\label{dOm}.
\ee
 
Most generally, we consider a two-zone IC scattering process, which can be reduced to the one-zone case straightforwardly. The IC scattering region does not overlap with the seed photon emission region, as shown in Figure \ref{fig1}. The incident photon flux in the IC scattering region is $F_{\nu_0,{\rm IC}}\sim cU_{\rm ph}/\nu_0\sim hcn_{\rm ph}$. The observed flux of the incident photons at earth is $F_{\nu_0}\sim(d/D)^2F_{\nu_0,{\rm IC}}$, where $D$ is the distance between the FRB emission region and the observer, and $d$ is the distance between the seed photon (FRB) emission region and the IC scattering region. Here $d\ll D$ is assumed.
Thus, the observed maximum flux of IC scattering is 
\be
F_{\nu,{\rm IC}}\sim (j_{\nu,{\rm IC}}L)\Omega_{0}\sim
\frac{4}{3}\sigma_{\rm T}n_eL\left(\frac{\Omega_{\rm IC}}{\Delta\Omega}\right)F_{\nu_0},
\ee
where $L$ is the line-of-sight length of the IC scattering region, $j_{\nu,{\rm IC}}L$ is the IC scattering intensity (an optically thin region is assumed), $\Omega_0$ is the solid angle of the IC scattering region opened to the observer, and $\Omega_{\rm IC}\sim(D/d)^2\Omega_0$ is the solid angle of the IC scattering region opened to the seed photon emission region, as shown in Figure \ref{fig1}. 
Defining the electron scattering optical depth as
\be
\tau_e=\sigma_{\rm T}n_eL, 
\ee
one finally has
\be
F_{\nu,{\rm IC}}\sim\frac{4\tau_e}{3}\left(\frac{\Omega_{\rm IC}}{\Delta\Omega}\right)F_{\nu_0}~~~~~{\rm at}~\nu\sim\gamma^2\nu_0.\label{ICs}
\ee
For the one-zone IC scattering model with non-beaming electrons, one has $\Omega_{\rm IC}\sim\Delta\Omega\sim4\pi$, so that $F_{\nu,{\rm IC}}\sim(4\tau_e/3)F_{\nu_0}$. 

We are interested in FOBs associated with FRBs.
If an FOB with $\nu_{\rm opt}\sim10^{15}~\unit{Hz}$ is generated via IC scattering of an FRB with $\nu_{\rm FRB}\sim10^9~\unit{Hz}$, the corresponding electron Lorentz factor would be 
\be
\gamma\sim\sqrt{\frac{\nu_{\rm opt}}{\nu_{\rm FRB}}}\simeq 10^3\left(\frac{\nu_{\rm opt}}{10^{15}~\unit{Hz}}\right)^{1/2}\left(\frac{\nu_{\rm FRB}}{10^{9}~\unit{Hz}}\right)^{-1/2}.\label{gamma}
\ee
Thus, the IC scattering flux mainly depends on the electron scattering optical depth with $\gamma\sim10^3$. In the following, we define $\eta_\gamma$ as the ratio between the number of electrons with $\gamma\sim10^3$ and the total electron number.

At last, the duration of FOB depends on the scale of the IC-scattering region. For one-zone IC-scattering processes, the optical duration is as same as that of an FRB, e.g., a few milliseconds; for two-zone IC-scattering processes, the FOB duration could be longer, since the scattering region could be much larger than the FRB emission region.

\subsection{One-zone: pulsar magnetosphere}\label{magnetosphere}

Most FRB models invoke emission within the magnetosphere of a strongly magnetized pulsar \citep[e.g.][]{kum17,yan18}. In this section, we consider a one-zone IC scattering model within such a framework, with the IC scattering occurring in the same region as the radio emission so that
$\Omega_{\rm IC}\sim 4\pi$.
If an FRB with $\nu_{\rm FRB}\sim1~\unit{GHz}$ is generated via coherent curvature radiation by electrons with $\gamma\sim10^3$, the curvature radius would be
\be
\rho=\frac{3}{4\pi}\frac{c\gamma^3}{\nu_{\rm FRB}}\simeq7\times10^{9}~\unit{cm}\left(\frac{\nu_{\rm opt}}{10^{15}~\unit{Hz}}\right)^{3/2}\left(\frac{\nu_{\rm FRB}}{10^{9}~\unit{Hz}}\right)^{-5/2},\nonumber\\\label{rho}
\ee
where Eq.(\ref{gamma}) has been used. 
In a pulsar magnetosphere, the number density of the electrons/positrons with $\gamma\sim10^3$ is $n_e\sim\eta_\gamma\mu_\pm n_{\rm GJ}$, where $\eta_\gamma$ is the fraction of the electrons/positrons with $\gamma\sim10^3$ in the total electron population, $\mu_\pm$ is the multiplicity resulting from the electron-positron pair cascade\footnote{For producing coherent curvature radiation with bunches, the contribution from electrons and positrons are cancelled out \citep{yan18} so that $\mu_\pm$ does not directly enter the problem to estimate the radio flux. However, for IC emission, both electrons and positrons contribute to the observed flux, so that $\mu_\pm$ is relevant in estimating the optical flux.}, and $n_{\rm GJ}=(r/R)^{-3}B/Pec$ with $r\sim\rho$ denoting the Goldreich-Julian density in the emission region, where $R=10^6~\unit{cm}$ is the neutron star radius. 

In the magnetosphere, the magnetic field line is curved and electrons are moving along the field line, thus a radio photon generated via curvature radiation will collide with other electrons at a point near the emission region of radio photon. 
First, we consider the IC scattering process from a single electron.
For the scattering photons within the aperture angle of $\sim1/\gamma$, the frequency of the scattering photon is \citep[e.g.][]{ghi13} 
\be
\nu\simeq\gamma^2(1-\beta\cos\theta)\nu_0,
\ee
where $\beta=\sqrt{1-1/\gamma^2}\sim1$ is the electron dimensionless velocity, $\theta$ is the collision angle defined as the angle between the electron momentum and the direction of the incident photon, and $\nu_0$ is the incident photon frequency. Therefore, once $\theta\gg0$, the scattering photon frequency is of the order of $\sim\gamma^2\nu_0$.

Since electrons move along the curved magnetic field lines, as the radio photons propagate in the curved field, the collision angle would gradually increase and eventually the scattering photon frequency would reach\footnote{There are many radio photons colliding with electrons with an collision angle less than $1/\gamma$. However, in this case, the scattered photons would have frequencies $\sim\nu_0$, and they would not reach the optical band. This is also the scenario of induced Compton scattering.} $\sim\gamma^2 \nu_0$. 
On the other hand, since the electrons are moving along with the curved field line, only the electrons at a part of the field line (with a length of $\sim\rho/\gamma$) can emit the observed scattering photon. Thus, the emission dimension of the IC-scattering region will be $L\sim\rho/\gamma$, from which the optical flux along the line of sight is emitted.

Since the electrons in the magnetosphere are beamed along curved magnetic field lines, the radiation beaming solid angle is taken as $\Delta\Omega\sim\pi/\gamma^2$ in Eq.(\ref{emission}) and Eq.(\ref{dOm}).
Finally, one has
\be
\frac{F_{\nu,{\rm IC}}}{F_{\nu,0}}&\sim&\frac{4\tau_e}{3}\left(\frac{\Omega_{\rm IC}}{\Delta\Omega}\right)\sim\frac{16}{3}\sigma_{\rm T}\eta_\gamma\mu_\pm n_{\rm GJ}\gamma\rho\simeq5\times10^{-5}\eta_\gamma\nonumber\\
&\times&\left(\frac{\mu_\pm}{10^3}\right)\left(\frac{B}{10^{14}~\unit{G}}\right)\left(\frac{P}{10~\unit{ms}}\right)^{-1}\label{ff1}
\ee
for $\gamma\simeq10^3$ and $\rho\simeq7\times10^9~\unit{cm}$, the typical values required by the FRB and FOB conditions 
(see Eq.(\ref{gamma}) and Eq.(\ref{rho})).

In order to constrain the optical flux, one needs to consider the possible range of pulsar parameters, e.g, magnetic field $B$ and period $P$. In the literature, a neutron star origin of FRBs has been discussed within the context of several energy powers: 1. the spindown power; 2. the magnetic power; 3. the gravitational power; and 4. the kinetic power. 

We first consider the case that FRBs are powered by pulsar spindown \citep[e.g.][]{met17}. In this case, the spindown luminosity $L_{\rm sd}$ needs be larger than the FRB luminosity $L_{\rm FRB}$ with a beaming fraction $f_b$, e.g. $L_{\rm sd}\gtrsim f_bL_{\rm FRB}$. The dipolar spindown luminosity $L_{\rm sd}$ can be written as
\be
L_{\rm sd}&=&L_{\rm sd,0}\left(1+\frac{t}{t_{\rm sd}}\right)^{-2}\simeq10^{43}~\unit{erg~s^{-1}}\nonumber\\
&\times&\left\{
\begin{aligned}
&0.4\left(\frac{B}{10^{14}~\unit{G}}\right)^{-2}\left(\frac{t}{1~\unit{yr}}\right)^{-2}&\text{for magnetar}~t\gg t_{\rm sd}\\
&\left(\frac{B}{10^{12}~\unit{G}}\right)^2\left(\frac{P_0}{1~\unit{ms}}\right)^{-4}&\text{for pulsar}~t\ll t_{\rm sd}\\
\end{aligned}
\right.\nonumber\\\label{spow}
\ee
where $L_{\rm sd,0}=I\Omega_0^2/2t_{\rm sd}\simeq10^{47}~\unit{erg~s^{-1}}(B/10^{14}~\unit{G})^{2}(P_0/1~\unit{ms})^{-4}$ is the initial spindown power, $t_{\rm sd}=3c^3I/B^2R^6\Omega_0^2=2\times10^5~\unit{s}(B/10^{14})^{-2}(P_0/1~\unit{ms})^2$ is the spindown timescale, $I=10^{45}~\unit{g~cm^2}$ is the moment of inertia of the neutron star, $\Omega_0=2\pi/P_0$ is the initial angular velocity of the neutron star. 

As shown in Eq.(\ref{spow}), FRBs could be emitted by magnetars with $B\simeq10^{14}~\unit{G}$ at $t\gg t_{\rm sd}$ or young pulsars with $B\gtrsim10^{12}~\unit{G}$ at $t\ll t_{\rm sd}$. We discuss the two cases in turn.
\begin{itemize}
\item Magnetars with $B\simeq10^{14}~\unit{G}$ at $t\gg t_{\rm sd}$: The requirement of $L_{\rm sd}\gtrsim f_b L_{\rm FRB}$ gives an age constraint 
\be
t\lesssim0.6~\unit{yr} \ f_b^{-1/2}\left(\frac{L_{\rm FRB}}{10^{43}~\unit{erg~s^{-1}}}\right)^{-1/2}\left(\frac{B}{10^{14}~\unit{G}}\right)^{-1},
\ee
and the corresponding spin period increases with time as
\be
P&\simeq&P_0\left(\frac{t}{t_{\rm sd}}\right)^{1/2}\simeq13~\unit{ms}\left(\frac{B}{10^{14}~\unit{G}}\right)\left(\frac{t}{1~\unit{yr}}\right)^{1/2}\nonumber\\
&\lesssim&10~\unit{ms}f_b^{-1/4}\left(\frac{L_{\rm FRB}}{10^{43}~\unit{erg~s^{-1}}}\right)^{-1/4}\left(\frac{B}{10^{14}~\unit{G}}\right)^{1/2}.\nonumber\\\label{period}
\ee
Note that in this case, $P(t)$ is independent of $P_0$.
According to Eq.(\ref{ff1}) and Eq.(\ref{period}), the optical flux is constrained by
\be
\frac{F_{\nu,{\rm IC}}}{F_{\nu,0}}&\gtrsim&5\times10^{-5}\eta_\gamma f_b^{1/4}\nonumber\\
&\times&\left(\frac{\mu_\pm}{10^3}\right)\left(\frac{B}{10^{14}~\unit{G}}\right)^{1/2}\left(\frac{L_{\rm FRB}}{10^{43}~\unit{erg~s^{-1}}}\right)^{1/4}.\nonumber\\\label{casea}
\ee
\item Young pulsars with $B\simeq10^{12}~\unit{G}$ at $t\ll t_{\rm sd}$: According to Eq.(\ref{spow}), the magnetic field needs to satisfy
\be
B\gtrsim10^{12}~\unit{G}f_b^{1/2}\left(\frac{L_{\rm FRB}}{10^{43}~\unit{erg~s^{-1}}}\right)^{1/2}\left(\frac{P_0}{1~\unit{ms}}\right)^{2},
\ee
and the period is about $P\sim1~\unit{ms}(P_0/1~\unit{ms})$. Thus, the optical flux is constrained by
\be
\frac{F_{\nu,{\rm IC}}}{F_{\nu,0}}&\gtrsim&5\times10^{-6}\eta_\gamma f_b^{1/2}\nonumber\\
&\times&\left(\frac{\mu_\pm}{10^3}\right)\left(\frac{L_{\rm FRB}}{10^{43}~\unit{erg~s^{-1}}}\right)^{1/2}\left(\frac{P_0}{1~\unit{ms}}\right).\label{caseb}
\ee

\end{itemize}

Next, we consider that FRBs are powered by dissipation of the pulsar magnetic fields \citep[e.g.][]{kum17}. 
In the FRB emission region, the magnetic energy density is $U_{\rm B}=B_e^2/8\pi$, where $B_e\sim B(r_e/R)^{-3}$ corresponds to the magnetic field strength at the emission radius $r_e\sim c\Delta t_{\rm FRB}$.
In order to produce FRBs, the magnetic energy density $U_{\rm B}$ should be larger than the FRB energy density, $U_{\rm FRB}=f_bL_{\rm FRB}/(4\pi r_e^2c)$, in the emission region. Thus, one has
\be
B&\gtrsim&\left(\frac{2c^3f_bL_{\rm FRB}\Delta t_{\rm FRB}^4}{R^6}\right)^{1/2}\nonumber\\
&\simeq&2\times10^{13}~\unit{G}f_b^{1/2}\left(\frac{L_{\rm FRB}}{10^{43}~\unit{erg~s^{-1}}}\right)^{1/2}
\ee
for $\Delta t_{\rm FRB}=1~\unit{ms}$.
According to Eq.(\ref{ff1}), for a pulsar with $B\gtrsim10^{13}~\unit{G}$ and $P\lesssim10~\unit{s}$, the optical flux is constrained by 
\be
\frac{F_{\nu,{\rm IC}}}{F_{\nu,0}}\gtrsim5\times10^{-9}\eta_\gamma f_b^{1/2}\left(\frac{\mu_\pm}{10^3}\right)\left(\frac{L_{\rm FRB}}{10^{43}~\unit{erg~s^{-1}}}\right)^{1/2}.\nonumber\\\label{casec}
\ee

Besides the spindown and magnetic powers, some FRB models invoke either the external gravitational energy power \citep[e.g.][]{gen15,dai16} or the external kinetic power \citep[e.g.][]{zha17b}. These models do not place significant requirements on the parameters of the neutron stars, so it is more reasonable to adopt the parameter distributions of the known pulsar population, e.g.
$B\sim(10^{10}-10^{15})~\unit{G}$, $P\sim(0.001-10)~\unit{s}$ and $\mu_\pm\sim(10^2-10^4)$. For such distributions, we get $F_{\nu,{\rm IC}}\sim5\times(10^{-13}-10^{-2})~\unit{Jy}$ for $F_{\nu_0}\sim1~\unit{Jy}$ and $\eta_\gamma\sim1$. 
At last, for one-zone IC-scattering process, the duration of IC-scattering photons is the same as that of the incident photons (the FRB itself).

\subsection{Two-zone: pulsar and nebula}\label{neb}

\begin{figure}
\centering
\includegraphics[angle=0,scale=0.32]{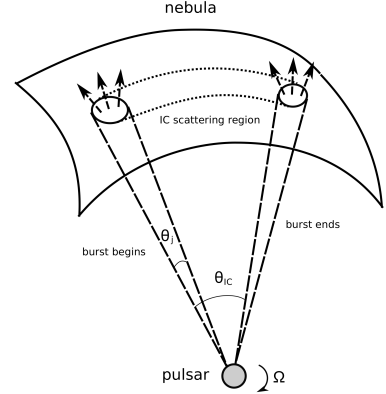}
\caption{A cartoon figure for a radio beam sweeping a surrounding nebula.}\label{fig2}
\end{figure}

Some authors suggested that at least some FRBs may be associated with a nebula such as a young supernova remnant (SNR) \citep{pir16,yan16,mur16,met17,bie17}.
Below we consider that the IC scattering region is in the nebula, and the incident photons i.e., the FRB, is generated via coherent radiation near a neutron star. Such a case belongs to a two-zone IC scattering model.
Different from the magnetospheric case discussed in Section \ref{magnetosphere}, the nebula may be isotropically opened to the central source and the electrons in the nebular emit isotropically, leading to $\Delta\Omega\sim4\pi$.
Thus, the optical flux may be estimated as
\be
F_{\nu,{\rm IC}}\sim\frac{\tau_e}{3\pi}\Omega_{\rm IC}F_{\nu_0}.\label{nebflux}
\ee
 
We consider that an FOB is generated by the IC-scattering process in a very young SNR with a relatively large electron column density.
For a young SNR, the electron number density is dominated by the ejecta component, which is \citep{yan17a}
\be
n_eL&\simeq&\frac{ML}{4\pi\mu_mm_pr^2L}\simeq2.6\times10^{4}~\unit{pc~cm^{-3}}\nonumber\\
&\times&\left(\frac{M}{M_\odot}\right)^2
\left(\frac{t}{1~\unit{yr}}\right)^{-2}\left(\frac{E}{10^{51}~\unit{erg}}\right)^{-1},\label{dmn}
\ee
where $n_e$ is the electron density in the SNR, $L$ is the SNR thickness, $\mu_m=1.2$ is the mean molecular weight for a solar composition in the SNR ejecta, $M$ is the SNR ejecta mass, $r\sim vt$ is the SNR radius, $v=\sqrt{2E/M}$ is the SNR ejecta velocity, $E$ is the SNR kinetic energy, and $t$ is the SNR age. $n_eL\sim{\rm DM}_{\rm SNR}$ just corresponds to the dispersion measure contributed by a freely expanding SNR \citep[e.g.][]{yan17a,pir18}, and the medium in the SNR is assumed to be ionized.

Near the center of the SNR, the FRB emission region may co-rotate with the magnetosphere, as shown in Figure \ref{fig2}. We define that the intrinsic duration of the incident photons (i.e. the intrinsic duration of the FRB) is $\Delta t_{\rm burst}$, and the burst ``jet'' beaming angle is $\theta_j$. 
Due to the possible rotation of the FRB source with period $P$,
the observed duration of an FRB satisfies 
\be
\Delta t_{\rm obs,0}\sim \min(\theta_j P/2\pi,\Delta t_{\rm burst}).
\ee 
Meanwhile, the IC scattering flux contributed by the isotropic electrons in the nebula is from an opening solid angle related to the neutron star, e.g., 
\be
\Omega_{\rm IC}\sim\theta_j\theta_{\rm IC}\sim\max(2\pi\Delta t_{\rm burst}\theta_j/P,\theta_j^2),
\ee 
where $\theta_{\rm IC}\sim\max(2\pi\Delta t_{\rm burst}/P,\theta_j)$ corresponds to a sweeping angle during the FRB intrinsic duration. 

According to Eq.(\ref{nebflux}), one finally has
\be
\frac{F_{\nu,{\rm IC}}}{F_{\nu_0}}\sim\frac{1}{3\pi}\sigma_{\rm T}\eta_\gamma n_eL\max\left(2\pi\Delta t_{\rm burst}\theta_jP^{-1},\theta_j^2\right).\label{ffneb}
\ee
At last, for the two-zone IC-scattering process discussed here, the scattering photons are from an extended region of the SNR, so that the duration of FOB could be longer than $\Delta t_{\rm burst}$, e.g.,
\be
\Delta t_{\rm obs, IC}&\sim&\frac{r\theta_{\rm IC}^2}{2c}\simeq5\times10^3~\unit{s}\left(\frac{t}{1~\unit{yr}}\right)\nonumber\\
&\times&\left(\frac{M}{M_\odot}\right)^{-1/2}
\left(\frac{E}{10^{51}~\unit{erg}}\right)^{1/2}\left(\frac{\theta_{\rm IC}}{0.1}\right)^2.\label{icobs}
\ee
Note that the duration of the IC scattering photons is longer than the typical exposure time of LSST/PTF, e.g. $T\sim\text{a few}\times10~\unit{s}$.

Based on the above result, e.g., Eq.(\ref{ffneb}), there are two typical cases:
\begin{itemize}
\item Case 1. the observed FRB duration $\Delta t_{\rm obs,0}\sim1~\unit{ms}$ is mainly determined by the time scale that the radio beam sweeps the line of sight, e.g., $\Delta t_{\rm obs,0}\sim\theta_j P/2\pi$. In this case, one has
\be
\frac{F_{\nu,{\rm IC}}}{F_{\nu_0}}&\sim&
\frac{\pi\sigma_{\rm T}\eta_\gamma n_eL\Delta t_{\rm burst}\Delta t_{\rm obs,0}}{3P^2}\nonumber\\
&\simeq&8.4\times10^{-9}\eta_\gamma\left(\frac{n_e}{1~\unit{cm^{-3}}}\right)\left(\frac{L}{1~\unit{pc}}\right)\left(\frac{\Delta t_{\rm burst}}{1~\unit{s}}\right)\nonumber\\
&\times&\left(\frac{\Delta t_{\rm obs,0}}{1~\unit{ms}}\right)\left(\frac{P}{1~\unit{s}}\right)^{-2}. \label{case1}
\ee
Note that since FRBs do not periodically repeat, the condition $\Delta t_{\rm burst}\lesssim P$ is required. 

\item Case 2. the observed FRB duration $\Delta t_{\rm obs,0}\sim1~\unit{ms}$ is mainly determined by the burst intrinsic duration, e.g., $\Delta t_{\rm obs,0}\sim\Delta t_{\rm burst}$. In this case, one has
\be
\frac{F_{\nu,{\rm IC}}}{F_{\nu_0}}&\sim&\frac{1}{3\pi}\sigma_{\rm T}\eta_\gamma n_eL\theta_j^2\nonumber\\
&\simeq&2.2\times10^{-9}\eta_\gamma\left(\frac{n_e}{1~\unit{cm^{-3}}}\right)\left(\frac{L}{1~\unit{pc}}\right)\left(\frac{\theta_j}{0.1}\right)^2. \label{case2}
\ee
\end{itemize}
In both cases, one has $F_{\nu,{\rm IC}}/F_{\nu_0}\sim10^{-9}\eta_\gamma(n_e/1~\unit{cm^{-3}})(L/1~\unit{pc})$ for typical parameters.
For a typical nebula, e.g., a normal SNR, with $n_e\sim1~\unit{cm^{-3}}$ and $L\sim1~\unit{pc}$, the IC scattering optical flux would be $F_{\nu,{\rm IC}}\sim10^{-9}~\unit{Jy}$ for a radio flux of $F_{\nu_0}\sim1~\unit{Jy}$ and $\eta_\gamma\sim1$, which is much fainter than the LSST sensitivity flux.

According to Eq.(\ref{dmn}), Eq.(\ref{case1}) and Eq.(\ref{case2}), one can write
\be
\frac{F_{\nu,{\rm IC}}}{F_{\nu,0}}&\simeq&10^{-4}\eta_\gamma\left(\frac{M}{M_\odot}\right)^2\left(\frac{t}{1~\unit{yr}}\right)^{-2}\left(\frac{E}{10^{51}~\unit{erg}}\right)^{-1}\nonumber\\
&\times&\left\{
\begin{aligned}
&2.2\left(\frac{\Delta t_{\rm burst}}{1~\unit{s}}\right)\left(\frac{\Delta t_{\rm obs,0}}{1~\unit{ms}}\right)\left(\frac{P}{1~\unit{s}}\right)^{-2}, & \text{for Case 1},\\
&0.56\left(\frac{\theta_j}{0.1}\right)^2, & \text{for Case 2}.
\end{aligned}
\right.
\nonumber\\\label{case0}
\ee

For a young SNR, the FRB emission may be subject to a large free-free opacity \citep[e.g.][]{lua14,mur16,met17} so that the FRB may not be detected. 
The free-free optical depth through the ejecta shell is 
\be
\tau_{\rm ff}&=&\alpha_{\rm ff}L\simeq(0.018T^{-3/2}Z^2n_en_i\nu^{-2}\bar g_{\rm ff})L\nonumber\\
&\simeq&3600~\left(\frac{T}{10^4~\unit{K}}\right)^{-3/2}\left(\frac{M}{M_\odot}\right)^{9/2}\left(\frac{E}{10^{51}~\unit{erg}}\right)^{-5/2}\nonumber\\
&\times&\left(\frac{t}{1~\unit{yr}}\right)^{-5}\left(\frac{\nu}{1~\unit{GHz}}\right)^{-2},
\ee
where $\bar g_{\rm ff}\sim1$ is the Gaunt factor, and $n_i$ and $n_e$ are the number densities of ions and electrons, respectively, $L\sim r\sim vt$ is the ejecta thickness. Here we assume that $n_e = n_i$ and $Z = 1$ for an ejecta shell with a fully ionized hydrogen-dominated composition. Fixing $\nu\sim1~\unit{GHz}$ and $T\sim10^4~\unit{K}$, for $\tau_{\rm ff}\lesssim1$, one gets the SNR age 
\be
t\gtrsim5~\unit{yr}\left(\frac{M}{M_\odot}\right)^{9/10}\left(\frac{E}{10^{51}~\unit{erg}}\right)^{-1/2}.\label{SNRage}
\ee
This result is consistent with \citet{met17}, who performed a detailed calculation for ionization and composition in the shell ejecta. Based on Eq.(\ref{case0}) and Eq.(\ref{SNRage}), the upper limit of the optical flux is
\be
\frac{F_{\nu,{\rm IC}}}{F_{\nu,0}}&\lesssim&10^{-6}\eta_\gamma\left(\frac{M}{M_\odot}\right)^{1/5}\nonumber\\
&\times&\left\{
\begin{aligned}
&8.8\left(\frac{\Delta t_{\rm burst}}{1~\unit{s}}\right)\left(\frac{\Delta t_{\rm obs,0}}{1~\unit{ms}}\right)\left(\frac{P}{1~\unit{s}}\right)^{-2}, & \text{for Case 1},\\
&2.2\left(\frac{\theta_j}{0.1}\right)^2, & \text{for Case 2}.
\end{aligned}
\right.
\nonumber\\\label{upsnr}
\ee
It is interesting that the constraint from the free-free absorption weakly depends on the SNR mass $M$ and is independent of the SNR kinetic energy $E$. Based on Eq.(\ref{upsnr}), we consider possible parameter distribution ranges for Case 1 and Case 2, respectively. For Case 1 with $\Delta t_{\rm burst}\lesssim P\sim(0.001-10)~\unit{s}$, one has $F_{\nu,{\rm IC}}\lesssim8.8\times10^{-3}~\unit{Jy}$ for $F_{\nu_0}\sim1~\unit{Jy}$ and $\eta_\gamma\sim1$. 
For Case 2 with $\theta_j\sim(0.01-1)$, one has $F_{\nu,{\rm IC}}\lesssim2.2\times10^{-4}~\unit{Jy}$ for $F_{\nu_0}\sim1~\unit{Jy}$ and $\eta_\gamma\sim1$. 
Since the optical duration is larger than the telescope exposure time, see Eq.(\ref{icobs}), for LSST with $F_{\nu\ast}\sim10^{-7}~\unit{Jy}$ for $\tau\gtrsim T\sim30~\unit{s}$, an FOB from a young SNR could be observable.
Again this conclusion is based on the assumption that most electrons have Lorentz factors around $\gamma\sim10^3$, so that $\eta_\gamma\sim1$ is satisfied. More generally, the observed optical flux would be much lower.

Recently, the MASTER-SAAO robotic telescope located at the South African Astronomical Observatory was pointed to FRB 181228A \citep{far18} 17723 seconds after the trigger time at 2018-12-28 18:44:13 UT. An upper limit up to 22.0 mag ($F_\nu\sim6\times10^{-6}~\unit{Jy}$)\citep{gor18} was derived. This time is too long to detect FOBs with a millisecond duration. However, the two-zone IC-scattering models invoking an SNR surrounding the FRB source is relevant. Based on Eq.(\ref{case0}), one can constrain the SNR parameters based on this follow-up observation. Assuming that $\eta_\gamma\sim1$, for an SNR with $M\simeq M_\odot$ and $E\simeq10^{51}~\unit{erg}$ surrounding an FRB source with age $t$, one can constrain $\Delta t_{\rm burst}\lesssim0.03~\unit{s}(P/1~\unit{s})^2(t/1~\unit{yr})^2$ for Case 1 and $\theta_j\lesssim0.03(t/1~\unit{yr})$ for Case 2.

\subsection{One zone: emission from masers in an outflow}\label{out}

Several authors have suggested various maser mechanisms to produce FRBs (e.g. \citealt{lyu14,ghi17,bel17,wax17,met19}, cf. \citealt{lu18}). 
Below we discuss the IC scattering processes in a maser outflow. In the maser outflow, the FRB emission is considered to be powered by the dissipation of free energy in the outflow at large distances \citep{sag02,wax17}.
When an expanding outflow drives a forward shock into the ambient plasma, a significant fraction of the kinetic energy of the outflow may be converted to internal energy within the shocked outflow plasma, then the internal energy may be radiated as an FRB via the synchrotron maser mechanism at the radius where the shell begins to decelerate \citep[e.g.][]{wax17,lon18,lu18,met19}. The maser emission is produced due to interaction between electromagnetic waves and energetic particles, leading to negative absorption.
 
In the following discussion, we consider that the IC scattering process from the maser electrons (with an inverted population distribution) that emit FRB photons at the same time. Such a scenario belongs to the one-zone IC scattering model. The distribution of the maser electrons in the outflow is assumed to be isotropic.
For an outflow moving along the line of sight with Lorentz factor $\Gamma$, 
an FRB is radiated from a shell of radius $r\sim2\Gamma^2c\Delta t_{\rm obs,0}$. 
The number density of the maser electrons in the comoving frame can be constrained by
$n_e'\sim L_{\rm iso}/4\pi r^2\Gamma^2\gamma m_ec^3$. where $\gamma\sim10^3$ corresponds to the local electron Lorentz factor in the shell comoving frame, which has been determined by the FRB frequency and the optical frequency\footnote{In the shell comoving frame, the FRB frequency is $\nu_{\rm FRB}'\sim\nu_{\rm FRB}/\Gamma$, and the IC-scattering frequency is $\nu_{\rm IC}'\sim\gamma^2\nu_{\rm FRB}'$. Therefore, in the observer frame, the IC-scattering frequency still satisfies the relationship $\nu_{\rm IC}\sim\Gamma\nu_{\rm IC}'\sim\gamma^2\nu_{\rm FRB}$.}, see Eq.(\ref{gamma}).
The optical depth of the IC scattering process is 
\be
\tau_e&\sim&\int\sigma_{\rm T}\eta_\gamma n_e(1-\beta)dr
\sim\frac{\sigma_{\rm T}\eta_\gamma n_e'r}{\Gamma}\sim\frac{\sigma_{\rm T}\eta_\gamma L_{\rm iso}}{8\pi \Gamma^5\gamma\Delta t_{\rm obs,0}m_ec^4}\nonumber\\
&\simeq&3.6\times10^{-8}\eta_\gamma\left(\frac{L_{\rm iso}}{10^{43}~\unit{erg~s^{-1}}}\right)\left(\frac{\Gamma}{100}\right)^{-5}\left(\frac{\Delta t_{\rm obs,0}}{1~\unit{ms}}\right)^{-1}\nonumber\\\label{taum}
\ee
for $\gamma=10^3$,
where $n_e=\Gamma n_e'$ is the electron density in the observer frame.
On the other hand, the induced Compton scattering in the maser outflow may suppress the observed FRB flux, and the corresponding Thompson optical depth can be constrained by \citep{lu18}
\be
\tau_{\rm T}\lesssim\frac{\Gamma^3\gamma^5m_ec^2}{kT_{\rm B}}\simeq6\times10^{-6}\left(\frac{\Gamma}{100}\right)^{3}\left(\frac{T_{\rm B}}{10^{36}~\unit{K}}\right)^{-1}\label{taue}
\ee
for $\gamma=10^3$, where $T_{\rm B}\simeq\Gamma^3T_{\rm B}'\sim10^{36}~\unit{K}$ is the FRB apparent brightness temperature, and $T_{\rm B}'$ is the brightness temperature in the shell comoving frame. Thus, according to Eq.(\ref{taum}) and Eq.(\ref{taue}), if the maser outflow is transparent for the FRB, 
the outflow Lorentz factor should satisfy $\Gamma\gtrsim50$.

For the one-zone IC scattering scenario, one has $\Omega_{\rm IC}\sim\Delta\Omega\sim4\pi$ and $F_{\nu,{\rm IC}}/F_{\nu_0}\sim4\tau_e/3$, so that
\be
\frac{F_{\nu,{\rm IC}}}{F_{\nu_0}}
&\simeq&1.6\times10^{-6}\eta_\gamma\nonumber\\
&\times&\left(\frac{L_{\rm iso}}{10^{43}~\unit{erg~s^{-1}}}\right)\left(\frac{\Gamma}{50}\right)^{-5}\left(\frac{\Delta t_{\rm obs,0}}{1~\unit{ms}}\right)^{-1}.
\ee
Therefore, for an outflow with $\Gamma\gtrsim50$, one has $F_{\nu,{\rm IC}}\lesssim1.6\times10^{-6}~\unit{Jy}$ for $F_{\nu_0}\sim1~\unit{Jy}$ and $\eta_\gamma\sim1$.
Again, the above conclusion still requires the most electrons have Lorentz factors around $\gamma\sim10^3$ to satisfy $\eta_\gamma\sim1$, which seems demanding. 
At last, since the above process is one-zone, the duration of IC-scattering photons is the same as that of the incident photons (FRBs).

\subsection{Two zone: IC scattering by galactic energetic electrons}

As discussed in Section \ref{ic}, the IC scattering flux mainly depends on the electron scattering optical depth. Another possibility to reach a reasonably large optical depth is that the scattering region has a large scale even though the electron number density is small. In the following, we consider the IC process by host galactic high-energy electrons off FRB seed photons.
According to the AMS observation \citep{agu14}, for Milky Way, the electron column density in the interstellar medium is about $\Phi_e\sim10^{-4}~\unit{cm^{-2}sr^{-1}s^{-1}}$ at $\sim1~\unit{GeV}$ ($\gamma\sim10^3$). Here we assume that the high-energy electron density in the FRB host galaxy is of the order of magnitude as that in Milky Way. Then the electron density is $\eta_\gamma n_e\sim4\pi \Phi_e/c\sim4\times10^{-14}~\unit{cm^{-3}}$ for electrons with $\gamma\sim10^3$. The optical depth is then about
\be
\tau_e&\sim& \sigma_{\rm T}\eta_\gamma n_eL\sim8\times10^{-16}\nonumber\\
&\times&\left(\frac{\Phi_e}{10^{-4}~\unit{cm^{-2}sr^{-1}s^{-1}}}\right)\left(\frac{L}{10~\unit{kpc}}\right).\nonumber\\
\ee 
Therefore, for a galaxy with a scale of $L\sim10~\unit{kpc}$, one has
$F_{\nu,{\rm IC}}\sim(4\tau_e/3)F_{\nu,0}\sim10^{-15}~\unit{Jy}$ for $F_{\nu,0}\sim1~\unit{Jy}$. For a host galaxy with less energetic electrons (e.g.,$\Phi_e<10^{-4}~\unit{cm^{-2}sr^{-1}s^{-1}}$) or a smaller scale (e.g.,$L\lesssim10~\unit{kpc}$), the corresponding optical flux is weaker. 
Therefore, the corresponding IC scattering optical flux would be extremely weak. 

\section{Optical emission by the FRB mechanism}\label{sec4}

In this section, we consider the optical emission component from the intrinsic mechanism of FRBs. Since FRBs show extremely high brightness temperatures, their radiation mechanisms must be coherent. For FRBs, two coherent mechanisms are often considered: radiation by bunches \citep{kat14,kum17,ghi17,kat18,yan18} and the maser mechanisms \citep{lyu14,bel17,ghi17b,wax17,lu18}. For optical emission component from the intrinsic mechanism of FRBs, its duration of the FOB should be the same as that of the FRB.

\subsection{Curvature radiation by bunches}

The curvature radiation by bunches is often suggested as the coherent radiation mechanism to explain the extreme brightness temperatures. When a charged particle moves along a curved trajectory, its perpendicular acceleration results in the so called ``curvature radiation''. Due to the extremely strong magnetic fields in the magnetosphere of the FRB engine, which is likely a neutron star with a rapid spin, electrons always move along the curved field lines and emit curvature radiation.

Recently, \citet{mac18} found that the mean spectral index of 23 fast radio bursts detected in a fly's-eye survey with the Australian SKA Pathfinder is $F_\nu\propto\nu^{\alpha}$ with $\alpha\sim-1.6$ \citep{mac18}. Theoretically, \citet{yan18} performed a detailed analysis about the coherent curvature radiation mechanism by bunches in a three-dimensional magnetic field geometry. They found that for three-dimensional bunches characterized by its length $L$, curvature radius $\rho$, bunch opening angle $\varphi$, given an electron energy distribution $N_e(\gamma)d\gamma\propto\gamma^{-p}d\gamma~{\rm for}~\gamma_1<\gamma<\gamma_2$, the curvature radiation spectra of the bunches are characterized by a multi-segment broken power law, with the break frequencies defined by $\nu_l=c/\pi L$, $\nu_\varphi=3c/2\pi\rho\varphi^3$ and $\nu_c=3c\gamma_1^3/4\pi\rho$.
Thus, the observed mean FRB spectra with $F_\nu\propto\nu^{-1.6}$ can be explained as one part of the multi-segment broken power law, and the corresponding spectral index depends on the electron distribution index $p$ and the order of $(\nu_l,\nu_\varphi,\nu_c)$ (see \citep[Figure 11 and Figure 12 in][]{yan18} for details). Within this theory, at higher frequencies, e.g. in the optical band, the coherence is suppressed due to the smaller wavelengths of the electromagnetic waves so that the spectrum is expected to be steeper. 
For all the cases, if $\nu>\nu_m=\max(\nu_l,\nu_\varphi,\nu_c)$ (which is the case for the optical band), one has $F_\nu\propto\nu^{-(2p+4)/3}$ \citep{yan18}. Very likely the curvature radiation spectrum appears $F_\nu\sim\nu^{-1.6}$ in the radio band but break to $F_\nu\propto\nu^{-(2p+4)/3}$ at higher frequencies.  
Due to the uncertainty of the break frequency, we estimate the optical flux to be between $F_{\nu,{\rm opt}}\simeq(\nu_{\rm opt}/\nu_{\rm radio})^{-1.6}F_{\nu,{\rm radio}}\sim2.5\times10^{-10}~\unit{Jy}(F_{\nu,{\rm radio}}/1~\unit{Jy})$ and $F_{\nu,{\rm opt}}\simeq(\nu_{\rm opt}/\nu_{\rm radio})^{-(2p+4)/3}F_{\nu,{\rm radio}}\lesssim10^{-16}~\unit{Jy}(F_{\nu,{\rm radio}}/1~\unit{Jy})$ with $p\gtrsim2$.
All these cases give an extremely low flux in the optical band, which is significantly below $F_{\nu{\ast}} \sim 0.01~\unit{Jy}$ for $\tau\sim1~\unit{ms}$.  

\subsection{Maser mechanism}

Maser emission is produced by inverted population of energetic particle, which results in negative absorption. Different from the normal absorption which causes the intensity decrease along a ray, the intensity actually increases when the maser condition is satisfied. For example, for an incident intensity of $I_{\nu,0}$, due to a negative absorption coefficient $\alpha_\nu$ at frequency $\nu$, the amplified intensity will be $I_\nu\sim I_{\nu,0}e^{-\tau_\nu}$, where $\tau_\nu=\alpha_\nu s<0$ is the optical depth and $s$ denotes the line-of-sight distance. At a coherent frequency $\nu$, the amplification is $\sim e^{|\tau_\nu|}$. 
Thus, for a flat incident spectrum, e.g., $F_{\nu}\propto\nu^0$, the flux ratio between incoherent optical emission and coherent radio emission would be $F_{\nu,{\rm opt}}/F_{\nu,{\rm radio}}\sim e^{-|\tau_{\nu}|}$. 

For negative synchrotron self-absorption discussed by \citet{wax17}, the optical depth is given by $|\tau|\sim20\epsilon_{B,-3}^{3/4}(E_{41}n_{-3}\Delta t_{-3})^{1/4}$, where $\epsilon_{B}=10^{-3}\epsilon_{B,-3}$ is the fractions of the internal energy carried by electrons and magnetic fields, $E=10^{41}E_{41}~\unit{erg}$ is the energy of a highly relativistic shells, $n=10^{-3}n_{-3}~\unit{cm^{-3}}$ is the ambient electron density, and $\Delta t=10^{-3}\Delta t_{-3}~\unit{s}$ is the FRB duration. Thus, even if one considers a weak maser mechanism  with $\epsilon_{B,-3}=1$, $E_{41}=1$, $n_{-3}=1$ and $\Delta t_{-3}=1$, one has the amplification $e^{|\tau_\nu|}\sim5\times10^8$ and the optical flux $F_{\nu,{\rm opt}}\sim2\times10^{-9}~\unit{Jy}(F_{\nu,{\rm radio}}/1~\unit{Jy})$ with the incident spectrum of $F_{\nu}\propto\nu^0$ assumed. For a stronger maser mechanism (e.g., $\epsilon_B\gg10^{-3}$ or $n\gg10^{-3}$) or a softer incident spectrum (e.g. $F_\nu\propto\nu^\alpha$ with $\alpha<0$), the predicted optical flux is even lower.

\section{Conclusions and Discussion}\label{sec5}

\begin{figure}
\centering
\includegraphics[angle=0,scale=0.35]{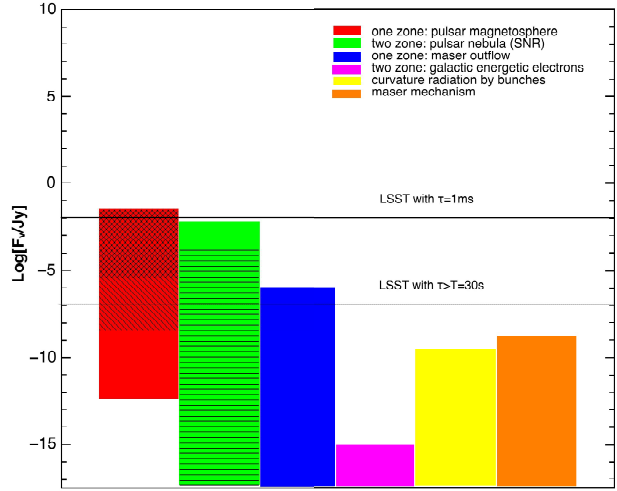}
\caption{The constraints on the optical flux of fast optical bursts from different models, including one-zone IC scattering in a pulsar magnetosphere (red), two-zone IC scattering in a pulsar wind nebula (green), one-zone IC scattering in a maser outflow (blue), two-zone IC scattering by galactic energetic electrons (purple), curvature radiation by bunches (yellow), and the maser mechanism (orange). 
In the red region (one-zone IC scattering in a pulsar magnetosphere), the cross shadow corresponds to the pulsar spindown-power mechanism, the diagonal shadow corresponds to the pulsar magnetic-power mechanism (overlapping with the spindown-power region), and the FRB beaming fraction is taken as $f_b=1$. 
In the green region (two-zone IC scattering in pulsar nebula), the entire green region corresponds to the FRB duration dominated by the radio beam sweeping time, and the horizontal shadow corresponds to the FRB duration dominated by the burst intrinsic duration.
The thick horizontal line denote the LSST sensitivity with a short integration time, i.e., $F_{\nu\ast}\sim0.01~\unit{Jy}$, with optical duration of $\tau=1~\unit{ms}$, and the
thin horizontal line denotes the LSST sensitivity with a long integration time, i.e.., $F_{\nu\ast}\sim10^{-7}~\unit{Jy}$, with optical duration of $\tau>T=30~\unit{s}$, where $T=30~\unit{s}$ corresponds to two exposures of 15 s of LSST.
One can see that these two IC scenarios (red and green) can lead to detectable FOBs only under extreme conditions.
For two-zone IC scattering in galactic energetic electrons (purple region), electrons have been considered to have Lorentz factors of $\gamma\sim10^3$ according to the AMS observations. For other IC-scattering models (red, green and blue regions), we assume that most IC-scatting electrons have Lorentz factors of $\gamma\sim10^3$, so that an optimistic efficiency $\eta_\gamma\sim1$ is assumed. 
The FRB flux is assumed to be $F_{\nu_0}\sim1~\unit{Jy}$.
}\label{fig3}
\end{figure}

We have discussed two astrophysical processes that may cause an FOB associated with an FRB: 1. IC scattering associated with an FRB; 2. the same FRB radiation mechanism extending to the optical band. The results are summarized in Figure \ref{fig3}.

We first discussed a list of possible FRB-associated IC scattering processes, including a one-zone model within the pulsar magnetosphere, a two-zone model invoking a pulsar nebula (e.g. SNR), a one-zone maser outflow model, as well as a two-zone model invoking galactic energetic electrons.
For one-zone IC-scattering processes, the optical duration is the same as that of an FRB. For two-zone IC-scattering processes, the optical duration could be longer, because the length scale of the scattering region could be much larger than that of the FRB emission region.

We note that in general
the optical flux from IC scattering processes in these scenarios is predicted to be very weak, which is much lower than $F_{\nu}\lesssim F_{\nu\ast}\sim0.01~\unit{Jy}$ (for optical duration with $\tau\sim1~\unit{ms}$) or $F_{\nu}\lesssim F_{\nu\ast}\sim10^{-7}~\unit{Jy}$ (for optical duration with $\tau\gtrsim T\sim \text{a few}\times10~\unit{s}$), the minimum detectable flux by LSST, as shown in Figure \ref{fig3}. 
Some extreme environments may provide an observable FOB, e.g, a neutron star with an extremely strong magnetic field ($B\sim10^{15}~\unit{G}$) and extremely fast rotation ($P\sim1~\unit{ms}$) (see Section \ref{magnetosphere}), or an FRB source surrounded by an extremely young SNR (see Section \ref{neb}).
Furthermore, even in such extreme environments, most electrons are required to have Lorentz factors around $\gamma\sim10^3$ in order to meet the requirement of $\eta_\gamma\sim1$. More generally, it is possible that $\eta_\gamma\ll1$, so that the number of the IC-scattering electrons  contributing to the optical flux would be much lower.
We next considered the optical emission due to the same mechanism that produces the FRB, including both the bunching coherent curvature radiation mechanism and the maser mechanism. Since optical emission is likely incoherent, the corresponding flux is also extremely weak, well below $F_{\nu}\lesssim F_{\nu\ast}\sim0.01~\unit{Jy}$, as shown in the yellow and orange regions in Figure \ref{fig3}. 

Besides FOBs, FRBs could be also associated with transient optical emission with a longer duration. A mechanism to produce optical emission that is weakly dependent on the FRB origin is the optical afterglow similar to that of GRBs. As studied by \citet{yi14}, such emission is much fainter than GRB afterglow due to the much smaller energy involved. The optical afterglow is bright enough for detection only when the condition is extreme, e.g. the FRB is very close and the total kinetic energy is large (i.e. a significant amount of energy does not show up in the radio band during the FRB phase). 

A bright optical transient may be expected for some specific models (e.g. \citet{pal18} for a living theory catalogue for FRBs). For example, within the double white dwarf merger model \citep{kas13}, a supernova may be associated with an FRB. The existence of such a bright supernova is excluded by the data for at least some FRBs, e.g. FRB 140514 \citep{pet15}. If some FRBs are associated with some violent catastrophic events such as double neutron star mergers \citep[e.g.][]{tot13,wan16,wan18} or GRBs \citep{zha14}, a bright optical component, either from the kilonova or the GRB afterglow, would be expected. 
The cosmic comb model \citep{zha17b} attributes an FRB to a sudden reconfiguration of the magnetosphere of a neutron star by a nearby ``astrophysical stream''. Within this scenario, FRBs may be associated with any type of violent processes, including supernovae/kilonovae, GRBs, AGN flares, and even tidal disruption events. A bright optical counterpart may be expected at least in some events\footnote{Within this scenario, the production of an FRB depends on the kinetic flux (not luminosity) received by the foreground neutron star. A low-luminosity event at a small distance from the neutron star can also produce a bright FRB. As a result, a bright optical counterpart is not always expected \citep{zha17b}.}.

Finally, in the above discussion, we take the projected limiting intrinsic flux of LSST with a few tens of seconds of exposure time to define the detectability of an FOB. However, thanks to the recent advancement of Complementary Metal Oxide Semiconductor (CMOS) technology, the exposure time can be improved to be shorter than a few hundreds of milliseconds. This would significantly improve the detection of transients with extremely short durations. On the other hand, even if the telescope sensitivity can be significantly improved in the future, the identification of FOBs with millisecond durations will be still a complicated task. If the FOB duration is much shorter than the exposure time, an FOB would be indistinguishable from some contaminating signals such as satellites and cosmic rays. Joint observations (between optical and radio telescopes, or between two optical telescopes at different sites) would be essential to screen FOBs from all kinds of noises.

\acknowledgments 
We thank an anonymous referee for helpful comments and suggestions, and Zhuo Li, Kai Wang, Jie-Shuang Wang, and Su Yao for helpful discussion.

\end{document}